\documentclass[twocolumn,showpacs]{revtex4}

\usepackage{graphics}

\begin{document}

\title{Photon-pair generation in random nonlinear
layered structures}

\author{Jan Pe\v{r}ina Jr.}
\affiliation{Joint Laboratory of Optics of Palack\'{y} University
and Institute of Physics of Academy of Sciences of the Czech
Republic, 17. listopadu 50A, 772 07 Olomouc, Czech Republic}
\email{perinaj@prfnw.upol.cz}
\author{Marco Centini}
\author{Concita Sibilia}
\author{Mario Bertolotti}
\affiliation{Dipartimento di Energetica, Universit\`{a} La
Sapienza di Roma, Via A. Scarpa 16, 00161 Roma, Italy}

\begin{abstract}
Nonlinearity and sharp transmission spectra of random 1D nonlinear
layered structures are combined together to produce photon pairs
with extremely narrow spectral bandwidths. Indistinguishable
photons in a pair are nearly unentangled. Also two-photon states
with coincident frequencies can be conveniently generated in these
structures if photon pairs generated into a certain range of
emission angles are superposed. If two photons are emitted into
two different resonant peaks, the ratio of their spectral
bandwidths may differ considerably from one and two photons remain
nearly unentangled.
\end{abstract}


\maketitle

\section{Introduction}

After the first successful experiment generating photon pairs in
the nonlinear process of spontaneous parametric down-conversion in
a nonlinear crystal and demonstration of its unusual temporal and
spectral properties more than 30 years ago \cite{Hong1987},
properties of the emitted photon pairs have been addressed in
detail in numerous investigations \cite{Mandel1995}. The key tool
in the theory of photon pairs has become a two-photon spectral
amplitude \cite{Keller1997,PerinaJr1999,DiGiuseppe1997,Grice1998}
that describes a photon pair in its full complexity. At the
beginning the effort has been concentrated on two-photon entangled
states with anti-correlated signal- and idler-field frequencies
that occur in usual nonlinear crystals. New sources able to
generate high photon fluxes have been discovered using, e.g.,
periodically-poled nonlinear crystals
\cite{Kuklewicz2005,Kitaeva2007}, four-wave mixing in nonlinear
structured fibers \cite{Li2005,Fulconis2005}, nonlinear planar
waveguides \cite{Tanzilli2002} or nonlinearities in cavities
\cite{Shapiro2000}. Chirped periodically-poled crystals opened the
door for the generation of signal and idler fields with extremely
wide spectral bandwidths \cite{Harris2007,Nasr2008}. Photon pairs
with wide spectral bandwidths can also be generated in specific
non-collinear geometries \cite{Carrasco2006}. Very sharp temporal
features are typical for such states that can be successfully
applied in metrology (see \cite{Carrasco2004} for quantum optical
coherence tomography). Later, even two-photon states with
coincident frequencies have been revealed. They can be emitted
provided that the extended phase matching conditions (phase
matching of the wave-vectors and group-velocity matching) are
fulfilled \cite{Giovannetti2002,Giovannetti2002a,Kuzucu2005}. Also
other suitable geometries for the generation of states with
coincident frequencies have been found. Nonlinear phase matching
for different frequencies at different angles of propagation of
the interacting fields can be conveniently used in this case
\cite{Torres2005,Torres2005a,Molina-Terriza2005}. Or wave-guiding
structures with transverse pumping and counter-propagating signal
and idler fields can be exploited
\cite{DeRossi2002,Booth2002,Walton2003,Ravaro2005,Lanco2006,Sciscione2006,Walton2004,PerinaJr2008}.
The last two approaches are quite flexible and allow the
generation of states having two-photon amplitudes with an
arbitrary orientation of main axes of their (approximately)
gaussian shape. These approaches emphasize the role of a
two-photon state at the boundary between the two mentioned cases.
This state is, a bit surprisingly, unentangled in frequencies and,
at the first sight, should not be very interesting. The opposed is
true \cite{URen2003,URen2005} due to the fact that the signal and
idler photons are completely indistinguishable and perfectly
synchronized in time for pulsed pumping. Non-collinear
configurations of bulk crystals of a suitable length and pump beam
with a suitable waist can also be used to generate such state
\cite{Grice2001,Carrasco2006}. These states with completely
indistinguishable photons are very useful in many
quantum-information protocols (e.g., in linear quantum computing)
that rely on polarization entanglement.

It has been shown that spectral entanglement affects polarization
entanglement and causes the lowering of polarization-entanglement
performance in quantum-information protocols. For example, the
role of spectral entanglement in polarization quantum
teleportation has been analyzed in detail in \cite{Humble2007}. We
note that this analysis is also appropriate for quantum repeaters,
quantum relays or quantum repeaters. If two photons are identical
and without any mutual entanglement, the polarization degrees of
freedom are completely separated from the spectral ones and in
principle the best possible performance of quantum information
protocols is guaranteed. In practice, mode mismatches both in
spectral and spatial domains may occur and degrade the
performance. It has been shown in \cite{Rohde2005} that Gaussian
distributed photons with large bandwidths represent states with
the best tolerance against errors. Under these conditions high
visibilities of interference patterns created by a simultaneous
detection of in principal many photons are expected.

As shown in this paper, nonlinear layered structures with randomly
positioned boundaries are a natural source of unentangled photon
pairs. If the down-converted photons are generated under identical
conditions (i.e. into the same transmission peaks) they are
indistinguishable and thus ideal for quantum-information
processing with polarization degrees of freedom. Moreover their
spectra are very narrow and temporal wave packets quite long and
so their use in experimental setups is fault tolerant against
unwanted temporal shifts \cite{Rohde2005}. Also because the
down-converted photons are generated into localized states with
high values of electric-field amplitudes higher photon-pair
generation rates are expected. This is important for protocols
exploiting several photon pairs. These states with very narrow
spectra can also be useful in entangled two-photon spectroscopy.
Superposition of photon pairs generated under different emission
angles is possible and two-photon states coincident in frequencies
can be engineered this way.

We note that the use of high electric-field amplitudes occurring
in localized states in random structures for the enhancement of
nonlinear processes has been studied in detail for second-harmonic
generation in \cite{Ochiai2005,Centini2006}.

The paper is organized as follows. The model of spontaneous
parametric down-conversion and formulas providing main physical
characteristics of photon pairs are presented in Sec.~II.
Properties of random 1D layered structures important for
photon-pair emission are studied in Sec.~III. A typical random
structure as a source of photon pairs is analyzed in Sec.~IV. A
scheme providing two-photon states coincident in frequencies is
analyzed in Sec.~V. Emission of a photon pair non-degenerate in
frequencies is investigated in Sec.~VI. Sec.~VII contains
Conclusions.

\section{Spontaneous parametric down-conversion and
characteristics of photon pairs}

Spontaneous parametric down-conversion in layered media can be
described by a nonlinear Hamiltonian $ \hat{H} $ \cite{Mandel1995}
in the formalism of quantum mechanics;
\begin{eqnarray}  
 \hat{H}(t) &=& \epsilon_0  \int_{\cal V} d{\bf r} \;  \nonumber \\
 & & \hspace{-10mm}  {\bf d}({\bf r}):
 \left[ {\bf E}_{p}^{(+)}({\bf r},t) \hat{\bf E}_{s}^{(-)}({\bf
r},t)
 \hat{\bf E}_{i}^{(-)}({\bf r},t) + {\rm h.c.} \right].
\label{1}
\end{eqnarray}
Vector properties of the nonlinear interaction are characterized
by a third-order tensor of nonlinear coefficients $ {\bf d} $
\cite{Boyd1994}. Symbol $ {\bf E}_{p}^{(+)} $ denotes a
positive-frequency pump-field electric-field amplitude, whereas
negative-frequency electric-field amplitude operators $ \hat{\bf
E}_{m}^{(-)} $ for $ m=s,i $ describe the signal and idler fields.
Symbol $ \epsilon_0 $ is permittivity of vacuum, $ {\cal V} $
interaction volume, $ {\rm h.c.} $ stands for a hermitian
conjugated term, and $ : $ means tensor multiplication with
respect to three indices. The electric-field amplitudes of the
pump, signal, and idler fields can be conveniently decomposed into
forward- and backward-propagating monochromatic plane waves when
describing scattering of the considered fields at boundaries of a
layered structure \cite{PerinaJr2006,Centini2005,Vamivakas2004}.
Vector properties of the electric-field amplitudes are taken into
account using decomposition into TE and TM modes. A detailed
theory has been developed in \cite{PerinaJr2006} and this theory
is used in our calculations. We note that a suitable choice of
vector and photonic properties can result in the generation of
photon pairs with a two-photon spectral amplitude anti-symmetric
in the signal- and idler-field frequencies \cite{PerinaJr2007}.
Some of the properties of photon pairs generated in layered
structures are common with these of photon pairs originating in
nonlinear crystal super-lattices composed of several nonlinear
homogeneous pieces \cite{URen2005a,URen2006}.

In further considerations, we assume fixed polarizations of the
signal and idler fields and restrict ourselves to a scalar model.
Then the perturbation solution of Schr\"{o}dinger equation for the
signal- and idler-fields wave-function results in the following
two-photon wave-function \cite{Keller1997,PerinaJr1999}:
\begin{eqnarray}   
 |\psi_{s,i}(t)\rangle &=&
  \int_{0}^{\infty} \, d\omega_s  \int_{0}^{\infty} \, d\omega_i \;
  \phi(\omega_s,\omega_i)
  \nonumber \\
 & & \mbox{} \hspace{-1.5cm}\times
  \hat{a}_{s}^{\dagger}(\omega_s)
  \hat{a}_{i}^{\dagger}(\omega_i) \exp(i\omega_s t) \exp(i
  \omega_i t)
  |{\rm vac} \rangle .
 \label{2}
\end{eqnarray}
The two-photon spectral amplitude $ \phi(\omega_s,\omega_i) $
giving a probability amplitude of generating a signal photon at
frequency $ \omega_s $ and its idler twin at frequency $ \omega_i
$ is determined for given angles of emission and polarizations of
the signal and idler photons. Creation operator $
\hat{a}_{m}^{\dagger}(\omega_m) $ adds one photon at frequency $
\omega_m $ into field $ m $ ($ i=s,i $) and $ |{\rm vac} \rangle $
means the vacuum state for the signal and idler fields. Both
photons can propagate either forward or backward outside the
structure as a consequence of scattering inside the structure.
Here we analyze in detail the case when both photons propagate
forward and note that similar behavior is found in the remaining
three cases. More details can be found in
\cite{PerinaJr2006,Centini2005}.

The two-photon spectral amplitude $ \phi $ can be decomposed into
the Schmidt dual basis with base functions $ f_{s,n} $ and $
f_{i,n} $ \cite{Law2000,Law2004,PerinaJr2008} in order to reveal
correlations (entanglement) between the signal and idler fields:
\begin{equation}   
 \phi(\omega_s,\omega_i) = \sum_{n=1}^{\infty} \lambda_n
 f_{s,n}(\omega_s) f_{i,n}(\omega_i) ;
\label{3}
\end{equation}
$ \lambda_n $ being coefficients of the decomposition. Entropy of
entanglement $ S $ defined as \cite{Law2000}
\begin{equation}  
 S = - \sum_{n=1}^{\infty} \lambda_n^2 \log_2 \lambda_n^2
\label{4}
\end{equation}
is a suitable quantitative measure of spectral entanglement
between the signal and idler fields. Symbol $ \log_2 $ stands for
logarithm with base 2. An additional measure of spectral
entanglement can be expressed in terms of the cooperativity
parameter $ K $ \cite{Law2004}:
\begin{equation}   
 K = \frac{1}{\sum_{n=1}^{\infty} \lambda_n^4 }.
\label{5}
\end{equation}

The two-photon spectral amplitude $ \phi(\omega_s,\omega_i) $
describes completely properties of photon pairs. Number $ N $ of
the generated photon pairs as well as signal-field energy spectrum
$ S_s(\omega_s) $ can simply be determined using the two-photon
spectral amplitude $ \phi $ \cite{PerinaJr2006}:
\begin{eqnarray}    
 N &=& \int_{0}^{\infty} d\omega_s \int_{0}^{\infty} d\omega_i
  |\phi(\omega_s,\omega_i)|^2 ,
\label{6} \\
 S_s(\omega_s)  &=& \hbar \omega_s \int_{0}^{\infty} d\omega_i
 |\phi(\omega_s,\omega_i)|^2.
\label{7}
\end{eqnarray}

Fourier transform $ \phi(t_s,t_i) $ of the spectral two-photon
amplitude $ \phi(\omega_s,\omega_i) $,
\begin{eqnarray}    
 \phi(t_s,t_i) &=& \frac{1}{2\pi}
  \int_{0}^{\infty} d\omega_s \; \int_{0}^{\infty} d\omega_i \;
  \sqrt{\frac{\omega_s\omega_i}{\omega_s^0\omega_i^0}}
  \phi(\omega_s,\omega_i)  \nonumber \\
  & & \mbox{} \times
 \exp(-i\omega_st_s)
 \exp(-i\omega_it_i),
\label{8}
\end{eqnarray}
is useful in determining temporal properties of photon pairs.
Symbol $ \omega_m^0 $ in Eq.~(\ref{8}) denotes the central
frequency of field $ m $, $ m=s,i $. This Fourier transform is
linearly proportional to the two-photon temporal amplitude $ {\cal
A} $ defined along the expression
\begin{eqnarray}    
  {\cal A}(t_s,t_i) &=&\langle {\rm vac} |
  \hat{E}^{(+)}_{s}(0,t_0+t_s) \hat{E}^{(+)}_{i}(0,t_0+t_i)
  |\psi_{s,i}(t_0)\rangle \nonumber \\
  & &
\label{9}
\end{eqnarray}
and giving a probability amplitude of detecting a signal photon at
time $ t_s $ together with its twin at time $ t_i $. We have
\cite{PerinaJr2006}:
\begin{equation}   
 {\cal A}(t_s,t_i) = \frac{\hbar \sqrt{\omega_s^0 \omega_i^0}}{4\pi\epsilon_0 c {\cal B}}
 \phi(t_s,t_i) ,
\label{10}
\end{equation}
where $ c $ is speed of light in vacuum and $ {\cal B} $
transverse area of the fields. Photon flux $ {\cal N } $ of, e.g.,
the signal field is determined using a simple formula (valid for
narrow spectra) \cite{PerinaJr2006}:
\begin{equation}    
 {\cal N}_{s}(t_s) = \hbar\omega_s^0
 \int_{-\infty}^{\infty} dt_i \; |\phi(t_s,t_i)|^2 .
\label{11}
\end{equation}

Direct measurement of temporal properties of photon pairs is
impossible because of short time scales needed and so temporal
properties may be detected only indirectly. Time duration of the
(pulsed) down-converted fields as well as entanglement time can be
experimentally addressed using Hong-Ou-Mandel interferometer.
Normalized coincidence-count rate $ R^{\rm HOM}_n $ in this
interferometer is derived as follows \cite{PerinaJr1999}:
\begin{eqnarray}   
 R^{\rm HOM}_n(\tau_l) &=& 1-\tilde{\rho}(\tau_l) , \\
 \tilde{\rho}(\tau_l) &=& \frac{1}{ R(0,0)}
 {\rm Re} \Biggl[ \int_{0}^{\infty} d\omega_s \, \int_{0}^{\infty} d\omega_i
 \; \omega_s \omega_i
 \nonumber \\
 & & \hspace{-15mm} \phi(\omega_s,\omega_i)
  \phi^{*}(\omega_i,\omega_s)
 \exp(i\omega_i\tau_l) \exp(-i\omega_s\tau_l) \Biggr] .
\label{13}
\end{eqnarray}
Normalization constant $ R(0,0) $ occurring in Eq.~(\ref{13}) is
given in Eq.~(\ref{15}) bellow. Relative time delay $ \tau_l $
between the signal and idler photons changes in the
interferometer.

A preferred direction of correlations between the signal- and
idler-field frequencies can be determined from the orientation of
coincidence-count interference fringes in Franson interferometer
\cite{Walton2003} that is characterized by normalized
coincidence-count rate $ R^{\rm F}_n $ in the form:
\begin{eqnarray}   
 R^{\rm F}_n(\tau_s,\tau_i) &=& \frac{1}{4} + \frac{1}{8R(0,0)} {\rm Re}
  \left\{ 2R(\tau_s,0) + 2R(0,\tau_i) \right. \nonumber \\
  & & \mbox{} \left.  + R(\tau_s,\tau_i) +
  R(\tau_s,-\tau_i) \right\}.
\label{14}
\end{eqnarray}
The function $ R $ in Eq.~(\ref{14}) is defined as follows:
\begin{eqnarray}   
 R(\tau_s,\tau_i) &=& \int_{0}^{\infty} d\omega_s \, \int_{0}^{\infty} d\omega_i
  \, \; \omega_s \omega_i |\phi(\omega_s,\omega_i)|^2  \nonumber \\
 & & \mbox{} \times \exp(i\omega_s\tau_s) \exp(i\omega_i\tau_i).
\label{15}
\end{eqnarray}
Time delay $ \tau_m $ ($ m=s,i $) corresponds to a relative phase
shift between two arms in the path of photon $ m $.

\section{Properties of random 1D layered structures}

We consider a 1D layered structure composed of two dielectrics
with mean layer optical thicknesses equal to $ \lambda_0/4 $ ($
\lambda_0 = 1\times 10^{-6} $~m is chosen). Such structure, for
example, can be fabricated by etching a crystal made of LiNbO$
\mbox{}_3 $ and filling free spaces with a suitable material (SiO$
\mbox{}_2 $). The optical axis of LiNbO$ \mbox{}_3 $ is parallel
to the planes of boundaries and coincides with the direction of
fields' polarizations that correspond to TE modes. This geometry
exploits the largest nonlinear coefficient of tensor $ {\bf d} $
($ {\bf d}_{zzz} $). The fields propagate as ordinary waves with
the dispersion formula $ n^2(\omega) = 4.9048 + 0.11768/(354.8084
- 0.0475\omega^2) - 9.6398 \omega^2 $ \cite{Dmitriev1991}. The
second material (SiO$ \mbox{}_2 $) is characterized by its index
of refraction $ n(\omega) = 1.45 $. A random structure is
generated along the following algorithm:
\begin{enumerate}
\item The number $ N_{\rm elem} $ of elementary layers of optical
 thicknesses $ \lambda_0/4 $ is fixed. Then the material of each
 elementary layer is randomly chosen.
\item At each boundary between two materials, an additional random
 shift of the boundary position governed by a gaussian distribution
 with variance $ \lambda_0/40 $ is introduced.
\end{enumerate}
The randomness given by a random choice of the material of each
layer is crucial for the observation of features typical for
random structures. The randomness introduced by gaussian shifts of
boundary positions is only additional and does not modify
considerably properties of a random structure. It also describes
errors occurring in the fabrication process. We note that the same
generation algorithm has been used in
\cite{Centini2006,Ochiai2005} where second-harmonic generation in
random structures has been addressed.

The process of spontaneous parametric down-conversion can be
efficient in these structures provided that the signal and idler
fields are generated into transmission peaks (corresponding to
spatially localized states) and the structure is at least
partially transparent for the pump field. This imposes
restrictions to the allowed optical lengths of suitable
structures. They have to have such lengths that the down-converted
fields (usually degenerate in frequencies) are localized whereas
there occurs no localization at the pump-field wavelength. This is
possible because the shorter the wavelength the larger the
localization length \cite{Anderson1958} for a given structure. We
assume the pump-field (signal-, idler-field) wavelength $
\lambda_p $ ($ \lambda_s $, $ \lambda_i $) in the vicinity of $
\lambda_0/2 $ ($ \lambda_0 $).

\begin{figure}    
 {\raisebox{4 cm}{a)} \hspace{0mm}
 \resizebox{0.7\hsize}{!}{\includegraphics{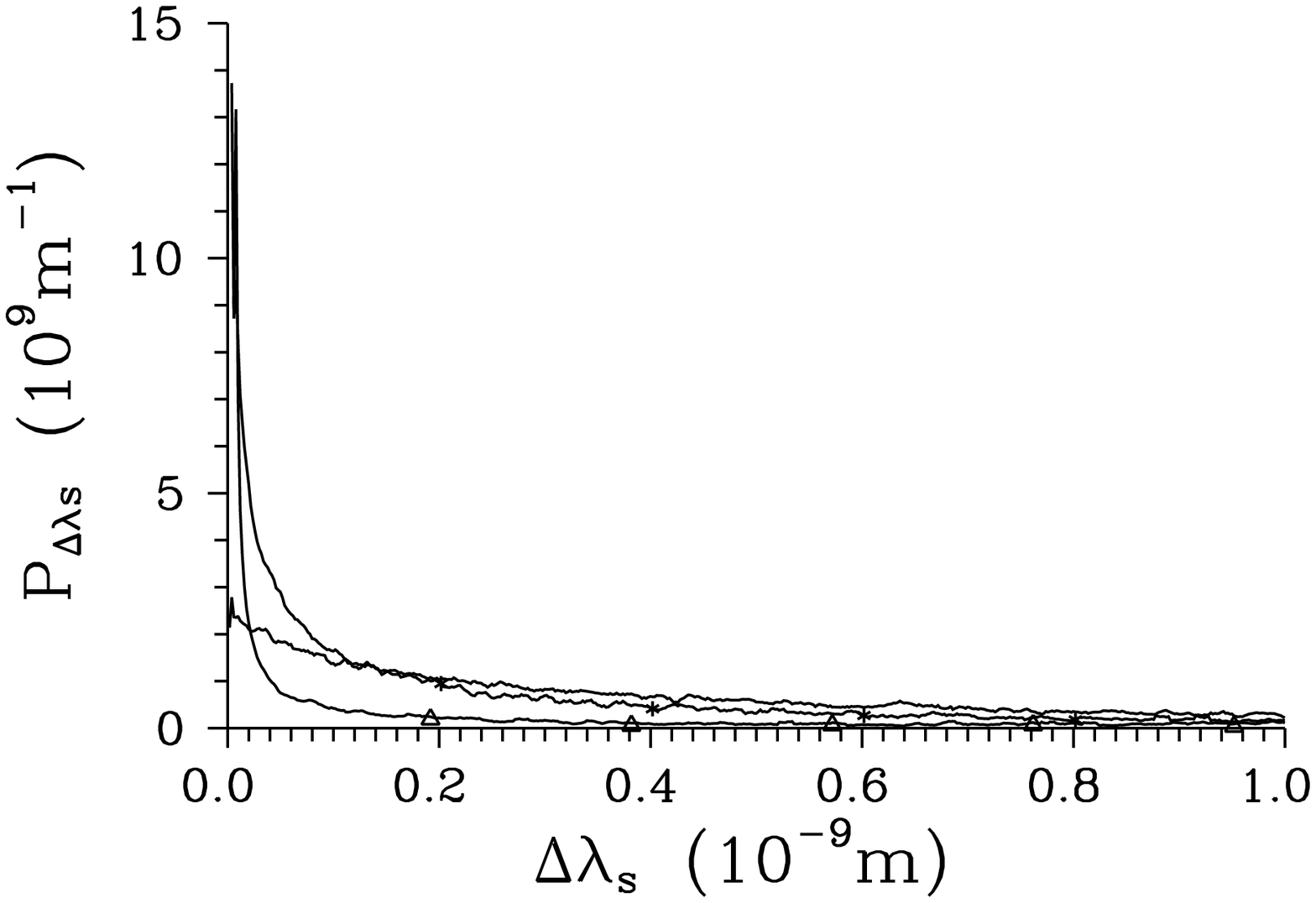}}

 \vspace{5mm}
 \hspace{0mm} \raisebox{2 cm}{b)} \hspace{5mm}
 \resizebox{0.7\hsize}{!}{\includegraphics{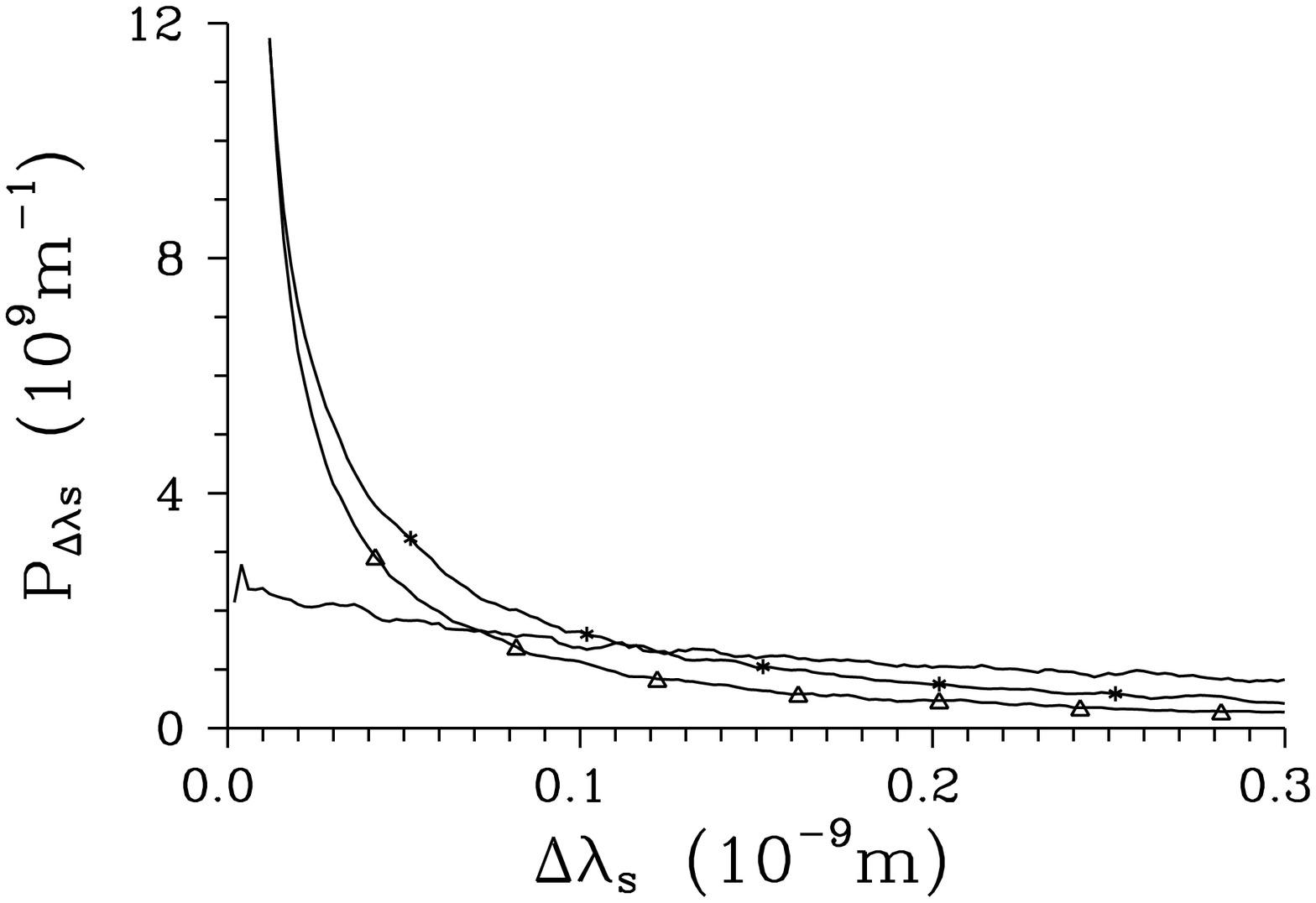}}
 }
 \vspace{2mm}

 \caption{Probability distribution $ P_{\Delta\lambda_s} $ of
 having a transmission peak with width (FWHM) $ \Delta\lambda_s $
 in the signal field at central wavelength near $ \lambda_0 $
 (a) for an ensemble with $ N_{\rm elem} = 250 $ (solid line),
 $ N_{\rm elem} = 500 $ (solid line with *), and $ N_{\rm elem} = 750 $
 (solid line with $ \triangle $) at $ \theta_s = 0 $~deg
 and (b) for angles $ \theta_s $ of signal-photon emission 0~deg (solid line),
 30~deg (solid line with *), and 60~deg (solid line with $
 \triangle $) using an ensemble with $ N_{\rm elem} = 250 $.}
\label{fig1}
\end{figure}

Numerical simulations for the considered materials and wavelengths
have revealed that the best suitable numbers $ N_{\rm elem} $ of
elementary layers lie between 200 and 400. In general, the greater
the contrast of indices of refraction of two materials the smaller
the number of needed layers. Widths of transmission peaks
corresponding to localized states in a random structure may vary
by several orders of magnitude. As an example, probability
distribution $ P_{\Delta\lambda_s} $ of widths of transmission
peaks for an ensemble of structures with $ N_{\rm elem} = 250 $
elementary layers (optical lengths of these structures lie around
$ 6 \times 10^{-5} $~m) is shown in Fig.~\ref{fig1}a. Comparison
with the distributions appropriate for $ N_{\rm elem} = 500 $ and
$ N_{\rm elem} = 750 $ in Fig.~\ref{fig1}a demonstrates that the
longer the structure the narrower transmission peaks can be
expected. Localization optical length at $ \lambda_0 $
\cite{Bertolotti2005,Centini2006} in the direction perpendicular
to the boundaries lies around  $ 22\times 10^{-6} $~m for the
considered structure lengths. As for the simulation, we have
randomly generated several tens of thousands structures in order
to have around $ 3\times 10^4 $ transmission peaks for the
statistical analysis. Transmission peaks can be easily
distinguished from their flat surroundings for which intensities
lower than one percent of the maximum peak intensity are typical.
Widths of intensity transmission peaks have been determined as
FWHM. Simulations have revealed that peaks with very small
intensity transmissions prevail, but there occur also peaks with
intensity transmissions close to one. Such peaks are useful for
the generation of photon pairs because of the presence of high
electric-field amplitudes inside the structure. Photon pairs can
be also generated under nonzero angles of emission. The greater
the radial angle $ \theta_s $ of signal-photon emission the
narrower the transmission peaks as documented in Fig.~\ref{fig1}b.
Also the localization lengths (projected into the direction
perpendicular to the boundaries) shorten with an increasing radial
angle $ \theta_s $ of signal-photon emission. Localization length
equal to $ 9.5 \times 10^{-6} $~m ($ 2.5 \times 10^{-6} $~m) has
been determined at the angle of signal-photon emission $ \theta_s
= 30 $~deg ($ 60 $~deg).

Fabrication of such structures is relatively easy due to allowed
high tolerances. A typical sample has several transmission peaks
at different angles $ \theta_s $ of emission for a given frequency
of the signal field. If we assume that also the idler field is
tuned into the same transmission peak the (normally incident) pump
field has to have twice the frequency corresponding to this peak.
In non-collinear geometry the transverse wave-vectors of the
signal and idler fields have to have the same magnitudes and
opposed signs in order to fulfill phase-matching conditions in the
transverse plane. If the angle $ \theta_s $ of signal-photon
emission increases a given transmission peak survives increasing
its central frequency $ \omega_s^0 $. Moreover the dependence of
central frequency $ \omega_s^0 $ on radial emission angle $
\theta_s $ can be considered to be linear in a certain interval of
angles $ \theta_s $. This property leads to wide tunability in
frequencies.

\section{Properties of the emitted photon pairs}

\begin{figure}    
 \resizebox{0.9\hsize}{!}{\includegraphics{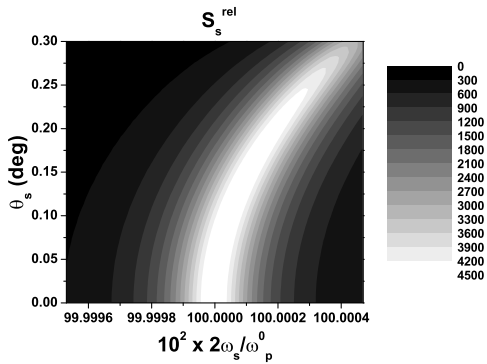}}
 \vspace{2mm}

 \caption{Contour plot of relative energy spectrum  $ S_s^{\rm rel} $
 of the signal field versus
 normalized signal-field frequency $ 2\omega_s/\omega_p^0 $ and angle
 $ \theta_s $ of signal-field emission. The relative spectrum  $ S_s^{\rm rel} $ is
 given by the ratio of actual spectrum of the structure and spectrum of
 a reference homogeneous structure containing the same amount of
 perfectly-phase-matched nonlinear material. The structure is pumped
 by a pulse 250~fs long.}
\label{fig2}
\end{figure}

We assume a normally incident pump beam forming a TE wave and
generation of the TE-polarized signal and idler fields into the
same transmission peak. Thus both emitted photons have the same
central frequencies and the central frequency of the pump field is
twice this frequency, i.e. the conservation law of energy is
fulfilled. If a structure is pumped by a broadband femtosecond
pulse, photon pairs are generated into a certain range of emission
angles. If a signal photon occurs at a given radial angle $
\theta_s $ its idler twin is emitted at its radial angle $
\theta_i = - \theta_s $ and wave-vectors of both photons and the
pump field lie in the same plane (i.e., their azimuthal angles $
\psi_s $ and $ \psi_i $ coincide) as a consequence of phase
matching conditions in the transverse plane. The central frequency
$ \omega_s^0 $ of a signal photon depends on the angle $ \theta_s
$ of signal-photon emission, as illustrated in Fig.~\ref{fig2}
showing the signal-field intensity spectrum $ S_s^{\rm ref} $.
Width of the spectrum $ S_s $ coincides with the width of
intensity transmission peak because the transmission peaks are
very narrow and so all frequencies inside them have nearly the
same conditions for the nonlinear process. This means that linear
properties of the photonic-band-gap structure have a dominant role
in the determination of spatial properties of the generated photon
pairs. Localization of the signal and idler fields inside the
considered photonic-band-gap structure leads to the enhancement of
photon-pair generation rate up to 5000 times (in the middle of
transmission peak) as can be deduced from Fig.~\ref{fig2}. This is
in accordance with the finding that the process of second-harmonic
generation can be enhanced by 3 or 4 orders of magnitude
\cite{Centini2006} in these structures. This enhancement is at the
expense of dramatic narrowing of the range of the allowed
frequencies of emitted photons. We estimate from 10 to $ 10^3 $
generated photon pairs into the whole emission cone per 100~mW of
the pump power depending on the width of transmission peak. The
wider the transmission peak the higher number of pairs is
expected. We note that extremely narrow down-converted fields can
be obtained also from parametric pumping of $ \mbox{}^{87} $Rb
atoms \cite{Kolchin2006}.

\begin{figure}    
 \resizebox{0.9\hsize}{!}{\includegraphics{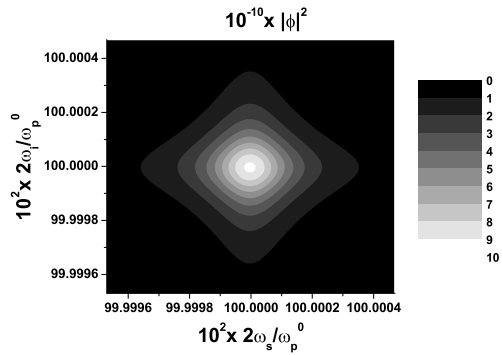}}
 \vspace{2mm}

 \caption{Contour plot of squared modulus $ |\phi|^2 $ of the two-photon spectral amplitude for
 the transmission peak at $ \theta_s = 0 $~deg. The structure is pumped
 by a pulse 250~fs long; $ |\phi|^2 $ is normalized such that
 $ 4 \int d\omega_s \int d\omega_i |\phi(\omega_s,\omega_i)|^2
 / (\omega_p^0)^2 = 1 $.}
\label{fig3}
\end{figure}

\begin{figure}    
 \resizebox{0.7\hsize}{!}{\includegraphics{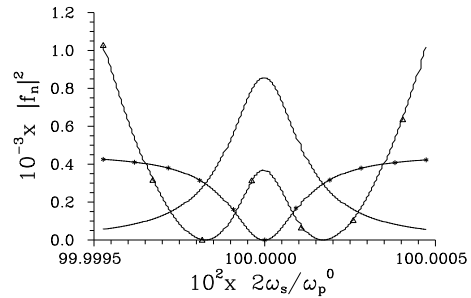}}
 \vspace{2mm}

 \caption{Squared modulus $ |f_n|^2 $ of mode functions with n=1 (solid line),
 2 (solid line with *), and 3 (solid line with $ \triangle $) of Schmidt decomposition
 of the two-photon spectral amplitude $ \phi $ in Fig.~\ref{fig3}; $
 f_{n,s}(\omega) = f_{n,i}(\omega) \equiv f_n(\omega) $ due to symmetry.
 Mode functions $ f_n $ are normalized
 such that $ 2\int d\omega |f_n(\omega)|^2 / \omega_p^0 = 1 $.}
\label{fig4}
\end{figure}

Two-photon spectral amplitudes $ \phi(\omega_s,\omega_i) $ for
different angles $ \theta_s $ of signal-photon emission are very
similar in their shape; they differ in their central frequencies.
We note that the central frequencies of the signal and idler
photons are the same ($ \omega_s^0 = \omega_i^0 $) for a given
angle $ \theta_s $ of emission. A typical shape of the two-photon
spectral amplitude $ \phi(\omega_s,\omega_i) $ resembling a cross
in its contour plot (see Fig.~\ref{fig3}) reflects the fact that
the signal and idler fields are nearly perfectly separable. This
is confirmed by Schmidt decomposition of the amplitude $
\phi(\omega_s,\omega_i) $ in which only the first mode is
important ($ \lambda_1 = 1.00 $). Entropy $ S $ of entanglement
for the two-photon spectral amplitude $ \phi $ in Fig.~\ref{fig3}
equals 0.00 and cooperativity parameter $ K $ is 1.00. The
spectral dependence of the first three mode functions in the
decomposition is shown in Fig.~\ref{fig4}.

A typical temporal two-photon amplitude $ {\cal A}(t_s,t_i) $
spreads over tens or hundreds of ps reflecting narrow frequency
spectra of the signal and idler fields. Its contour plot resembles
a droplet \cite{PerinaJr2006} that originates in the zig-zag
movement of the emitted photons inside the structure that delays
the occurrence time of photons at the output plane of the
structure. Both photons have the same wave-packets due to
identical emission conditions as can be verified in Hong-Ou-Mandel
interferometer showing the visibility equal to one.

Assuming the normally incident pump beam the signal-field
intensity profile at a given frequency in the transverse plane is
nonzero around a circle due to the rotational symmetry of the
photonic-band-gap structure around the pump-beam propagation
direction. This symmetry assumes an appropriate rotation of
polarization base vectors. However, values of intensity change
around the circle depending on fields' polarizations (defined by
analyzers in front of detectors) and properties of nonlinear
tensor $ {\bf d} $.

\begin{figure}    
 \resizebox{0.7\hsize}{!}{\includegraphics{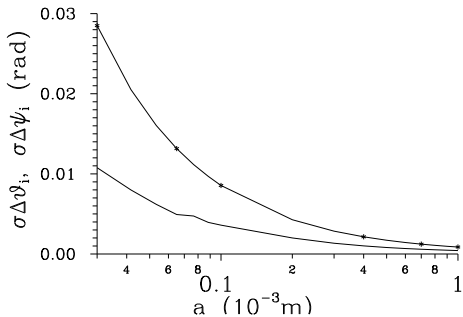}}
 \vspace{2mm}

 \caption{Spread of the correlation area in radial ($ \sigma\Delta\theta_i $, solid line)
  and azimuthal ($ \sigma\Delta \psi_i $, solid line with *) directions of idler-photon emission as a
  function of pump-beam diameter $ a $. The transmission peak in the structure
  with 250 elementary layers occurs at $ \theta_s^0 = 26.3 $~deg, i.e. the central radial
  idler-photon emission angle $ \theta_i^0 $ equals - 26.3~deg. Logarithmic scale on the
  $ x $ axis is used.}
\label{fig5}
\end{figure}

Correlations between the signal and idler fields in the transverse
plane are characterized by a correlation area that is defined by
the probability of emitting an idler photon in a given direction
[defined by angular declinations $ \Delta\theta_i $ (radial
direction) and $ \Delta \psi_i $ (azimuthal direction) from the
ideal direction of emission] provided that the direction of
signal-photon emission is fixed. Correlation area is in general an
ellipse with typical lengths in radial ($ \sigma\Delta\theta_i $)
and azimuthal ($ \sigma\Delta \psi_i $) directions. These angles
are very small for plane-wave pumping, typically of the order of $
10^{-3} - 10^{-4} $~rad. However, focusing of the pump beam can
increase their values considerably as shown in Fig.~\ref{fig5},
where the pump-beam diameter $ a $ varies from 30~$ \mu $m up to
1~mm. Spread of the correlation area in radial direction is
smaller compared to that in azimuthal direction, because emission
of a photon is more restricted in radial direction (for geometry
reasons) by narrow bands of the photonic structure. Release of
strict phase-matching conditions in the transverse plane caused by
a focused pump beam (with a circular spot) affects radial and
azimuthal angles of emission in the same way.

We note that spreading of the signal- and idler-field intensity
profiles caused by pump-beam focusing has been experimentally
observed in \cite{Zhao2008} for a type-II bulk crystal.

\section{Two-photon states coincident in frequencies}

Superposition of photon pairs with signal photons emitted under
different radial emission angles $ \theta_s $ results in states
coincident in frequencies. This spectral beam combining of fields
from different spatial modes and with different spectral
compositions can be achieved using an optical dispersion element
like a grating \cite{Augst2003}. This technique has already been
applied in the construction of a source of photon pairs that uses
achromatic phase matching and spatial decomposition of the pump
beam \cite{Torres2005,Torres2005a}.

\begin{figure}    
 \resizebox{0.9\hsize}{!}{\includegraphics{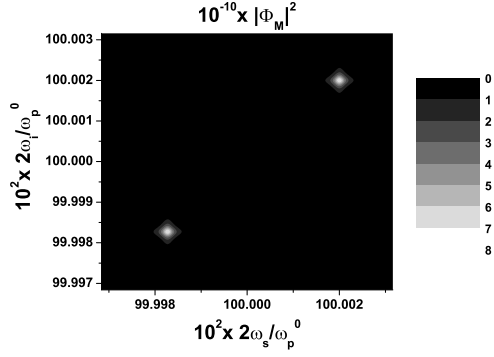}}
 \vspace{2mm}

 \caption{Contour plot of squared modulus $ |\Phi_M|^2 $ of the two-photon spectral
 amplitude created by superposing photon-pair amplitudes from $ M=2 $
 pinholes. The structure is pumped by a pulse 250~fs long; $ |\Phi_M|^2 $
 is normalized such that $ 4 \int d\omega_s \int d\omega_i |\Phi_M(\omega_s,\omega_i)|^2
 / (\omega_p^0)^2 = 1 $.}
\label{fig6}
\end{figure}

\begin{figure}    
 \resizebox{0.7\hsize}{!}{\includegraphics{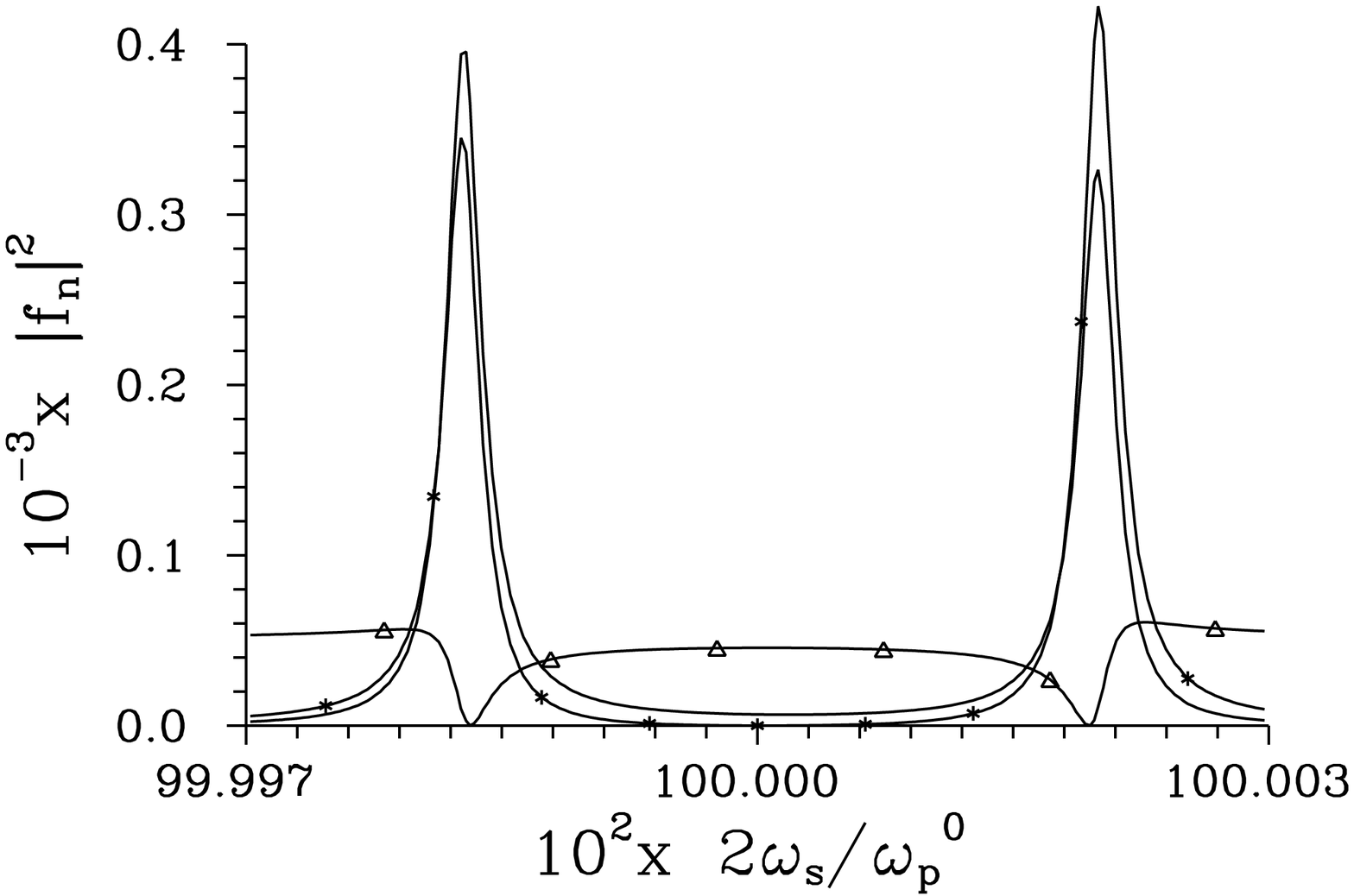}}
 \vspace{2mm}

 \caption{Squared modulus $ |f_n|^2 $ of mode functions with n=1 (solid line),
 2 (solid line with *), and 3 (solid line with $ \triangle $) of Schmidt decomposition
 of the two-photon spectral amplitude $ \Phi_M $ in Fig.~\ref{fig6}; $
 f_{n,s}(\omega) = f_{n,i}(\omega) \equiv f_n(\omega) $ due to symmetry.
 Mode functions $ f_n $ are normalized
 such that $ 2\int d\omega |f_n(\omega)|^2 / \omega_p^0 = 1 $.}
\label{fig7}
\end{figure}

There is approximately a linear dependence between the central
frequencies $ \omega_s^0 = \omega_i^0 $ and radial angle $
\theta_s $ of signal-field emission for sufficiently large values
of $ \theta_s $ assuming a normally incident pump beam. The
resultant two-photon spectral amplitude $
\Phi_M(\omega_s,\omega_i) $ after superposing photon pairs from $
M $ equidistantly positioned pinholes present both in the signal
and idler beams can be approximately expressed as:
\begin{equation}   
 \Phi_M(\omega_s,\omega_i) = \sum_{n=0}^{M-1} \exp(i\varphi n)
 \phi(\omega_s + n\Delta\omega,\omega_i+ n\Delta\omega) ,
\label{16}
\end{equation}
where $ \Delta\omega $ is the difference between the signal-field
central frequencies of fields originating in adjacent pinholes.
Phase $ \varphi $ determines the difference in phases of two
two-photon spectral amplitudes at the central signal- and
idler-field frequencies coming from adjacent pinholes. The
two-photon spectral amplitude $ \phi (\omega_s,\omega_i) $ in
Eq.~(\ref{16}) belongs to a two-photon state coming from the first
pinhole. Schmidt decomposition of the two-photon spectral
amplitude $ \Phi_M $ describing photon pairs coming from $ M $
pinholes shows that there are nearly $ M $ independent modes
[determined by the value of cooperativity parameter $ K $
introduced in Eq.~(\ref{5})]. These modes are collective modes,
i.e. they have non-negligible values in areas of frequencies
associated with every pinhole (see Figs.~\ref{fig6} and \ref{fig7}
for $ M=2 $ pinholes). Such states are perspective for the
implementation of various quantum-information protocols.

\begin{figure}    
 {\raisebox{4 cm}{a)} \hspace{0mm}
 \resizebox{0.9\hsize}{!}{\includegraphics{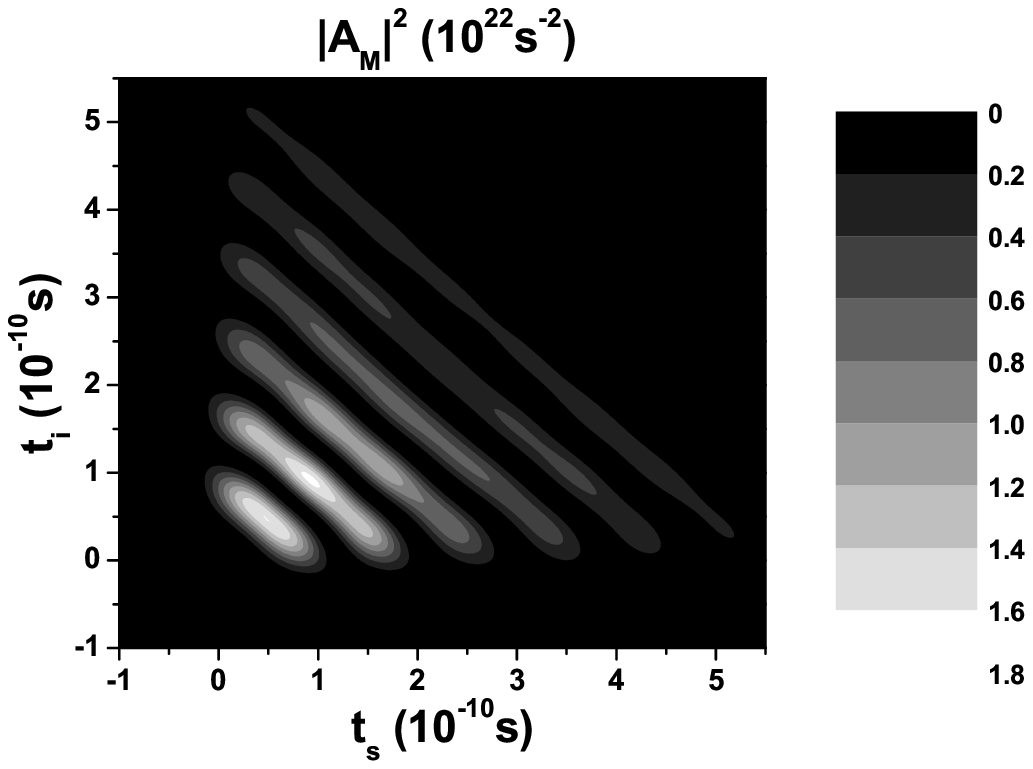}}

 \vspace{5mm}
 \hspace{0mm} \raisebox{4 cm}{b)} \hspace{0mm}
 \resizebox{0.9\hsize}{!}{\includegraphics{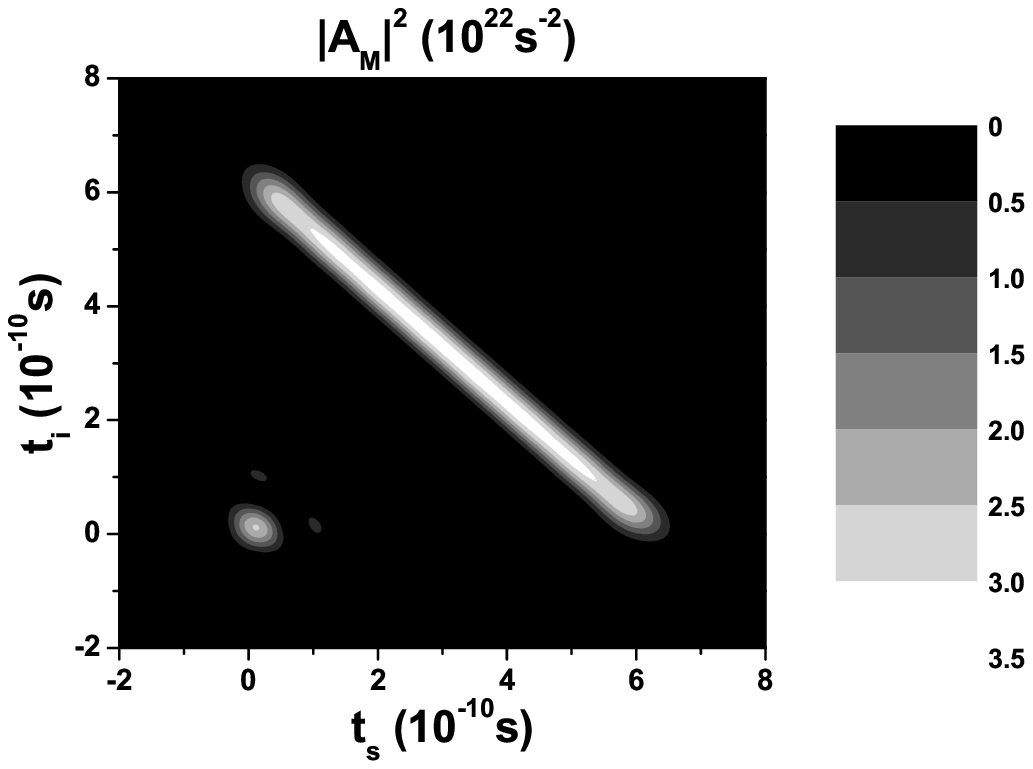}}
 }
 \vspace{2mm}

 \caption{Contour plot of squared modulus $ |{\cal A}_M|^2 $
 of the temporal two-photon amplitude for (a)
 $ M=2 $ and (b) $ M=8 $. The corresponding two-photon
 spectral amplitude $ \Phi_M $ for $ M=2 $ is plotted in
 Fig.~\ref{fig6}. Positions of the edge pinholes for $ M=8 $
 coincide with those for $ M=2 $ and the remaining six pinholes
 are equidistantly distributed in-between them. Normalization is such that
 $ \int dt_s \int dt_i |{\cal A}_M(t_s,t_i)|^2 = 1 $.}
\label{fig8}
\end{figure}

\begin{figure}    
 {\raisebox{4 cm}{a)} \hspace{0mm}
 \resizebox{0.9\hsize}{!}{\includegraphics{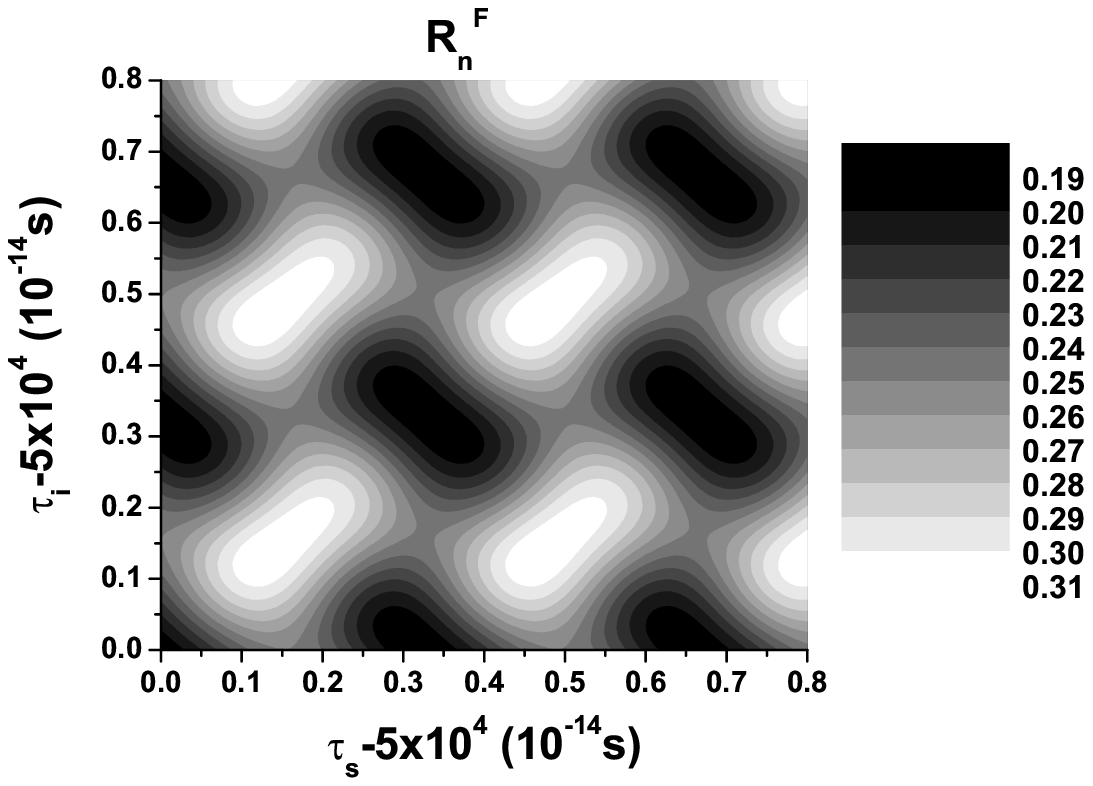}}

 \vspace{5mm}
 \hspace{0mm} \raisebox{4 cm}{b)} \hspace{0mm}
 \resizebox{0.9\hsize}{!}{\includegraphics{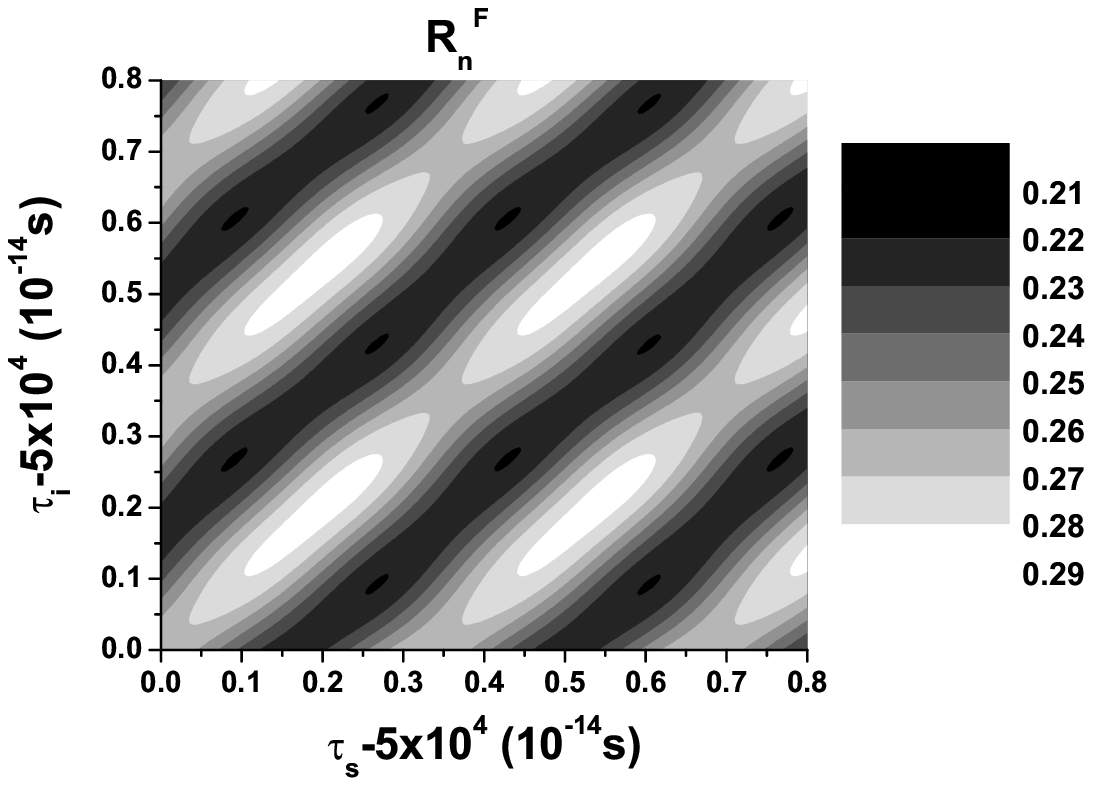}}
 }
 \vspace{2mm}

 \caption{Contour plot of normalized coincidence-count rate $ R_n^{\rm F} $ in Franson
  interferometer as a function of signal- and idler-field delays $ \tau_s $
  and $ \tau_i $ for (a) $ M=1 $ and (b) $ M=8 $ pinholes. Positions of pinholes
  are defined in caption to Fig.~\ref{fig8}.}
\label{fig9}
\end{figure}

The Fourier transform $ \Phi_M(t_s,t_i) $ defined in Eq.~(\ref{8})
can be approximately expressed as follows:
\begin{equation}   
 \Phi_M(t_s,t_i) \approx \sum_{n=0}^{M-1} \exp[i(\varphi -
  \Delta\omega(t_s+t_i))n] \phi(t_s,t_i) ,
\label{17}
\end{equation}
where
\begin{eqnarray}   
 \phi(t_s,t_i) &=& \frac{1}{2\pi} \int_{0}^{\infty} d\omega_s \int_{0}^{\infty}
  d\omega_i \phi(\omega_s,\omega_i) \exp(i\omega_s t_s) \nonumber
  \\
 & & \mbox{} \times  \exp(i\omega_i t_i)
\label{18}
\end{eqnarray}
means the Fourier transform of a two-photon amplitude associated
with one pinhole. The expression in Eq.~(\ref{17}) indicates that
interference of photon-pair amplitudes originating in different
pinholes creates interference fringes in the sum $ t_s + t_i $ of
the occurrence times of signal and idler photons. This is caused
by positive correlations in the signal- and idler-field
frequencies. The analysis of the sum in Eq.~(\ref{17}) shows that
the greater the number $ M $ of pinholes, the narrower the range
of allowed values of the sum $ t_s + t_i $ [$ \lim_{M\rightarrow
\infty} \sum_{n=0}^{M-1} \exp(ixn) \approx \delta(x) $].
Comparison of shapes of the two-photon temporal amplitudes $ {\cal
A}_M(t_s,t_i) $ derived in Eq.~(\ref{10}) for $ M=2 $ and $ M=8 $
pinholes shown in Fig.~\ref{fig8} reveals this tendency for
localization in time domain.
These features originating in positive correlations of the signal-
and idler-field frequencies are experimentally accessible by
measuring the pattern of coincidence-count rate $ R_n^{\rm F} $ in
Franson interferometer for sufficiently large values of signal-
and idler-photon delays $ \tau_s $ and $ \tau_i $. A (nearly
separable) two-photon state from one pinhole creates a chessboard
tilted by 45~degrees (see Fig.~\ref{fig9}a). If amplitudes from
several pinholes are included typical fringes oriented at 45
degrees become visible in coincidence-count patterns given by
rate $ R_n^{\rm F} $ (compare Figs.~\ref{fig9}a and \ref{fig9}b).
The greater the number $ M $ of pinholes, the better the fringes
are formed.

\begin{figure}    
 \resizebox{0.9\hsize}{!}{\includegraphics{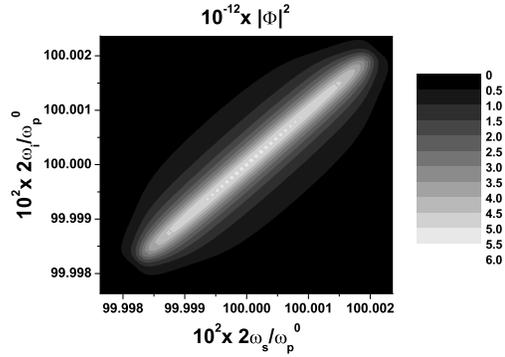}}
 \vspace{2mm}

 \caption{Contour plot of squared modulus $ |\Phi|^2 $ of the two-photon spectral
 amplitude created by superposition of photon-pair amplitudes from a certain range
 of signal-field (and idler-field) emission angles $ \theta_s $ ($ \theta_i $) with
 a suitable phase compensation. The structure is pumped by a pulse 250~fs long;
 $ |\Phi|^2 $ is normalized according to the condition
 $ 4 \int d\omega_s \int d\omega_i |\Phi(\omega_s,\omega_i)|^2
 / (\omega_p^0)^2 = 1 $.}
\label{fig10}
\end{figure}

\begin{figure}    
 \resizebox{0.7\hsize}{!}{\includegraphics{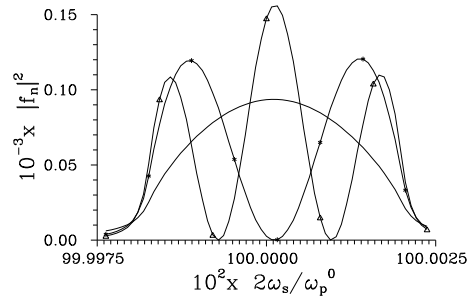}}
 \vspace{2mm}

 \caption{Squared modulus $ |f_n|^2 $ of mode functions with n=1 (solid line),
 2 (solid line with *), and 3 (solid line with $ \triangle $) of Schmidt decomposition
 of the two-photon spectral amplitude $ \Phi $ in Fig.~\ref{fig10}; $
 f_{n,s}(\omega) = f_{n,i}(\omega) \equiv f_n(\omega) $ due to symmetry.
 Mode functions $ f_n $ are normalized
 such that $ 2\int d\omega |f_n(\omega)|^2 / \omega_p^0 = 1 $.}
\label{fig11}
\end{figure}

Superposition of photon-pair amplitudes (with a suitable phase
compensation) from a given range of signal-field emission angles $
\theta_s $ (and the corresponding range of idler-field emission
angles $ \theta_i $) as defined by rectangular apertures gives a
two-photon spectral amplitude $ \Phi $ having a cigar shape with
coincident frequencies (see Fig.~\ref{fig10}). The phase
compensation requires the introduction of an additional phase
shift depending on the radial emission angle $ \theta_s $ and can
be accomplished, e.g., by a wedge-shaped prism. Schmidt
decomposition of the two-photon amplitude $ \Phi $ shows that
several modes are non-negligibly populated ($ \lambda_1 = 0.78 $,
$ \lambda_2 = 0.13 $, and $ \lambda_3 = 0.04 $). Typical profiles
of mode functions $ f_n $ for this configuration are plotted in
Fig.~\ref{fig11}. The first and most populated mode extends over
all frequencies whereas oscillations are characteristic for the
other modes. The larger the spectral width of the signal and idler
fields the larger the cooperativity parameter $ K $. This means
that the number of effective independent modes can be easily
changed just by changing the height of rectangular apertures. This
makes this source of photon pairs extraordinarily useful. Both
emitted photons are perfectly indistinguishable providing
visibility equal to one in Hong-Ou-Mandel interferometer. Positive
correlation in the signal- and idler-field frequencies gives
coincidence-count interference fringes in Franson interferometer
tilted by 45 degrees for sufficiently large values of the signal-
and idler-field delays $ \tau_s $ and $ \tau_i $.

\section{Spectrally non-degenerate emission of photon pairs}

The simplest way for the observation of spectrally non-degenerate
emission of a photon pair is to consider collinear geometry and
exploit a random structure with two different transmission peaks.
Pumping frequency is then given as the sum of central frequencies
of these transmission peaks into which signal and idler photons
are emitted. The generation of a suitable structure using the
algorithm and geometry presented in Sec.~III in this case is by an
order of magnitude more difficult compared to that providing
photon pairs degenerated in frequencies. Structures having two
peaks with considerably different bandwidths of the transmission
peaks are especially interesting. Spectral bandwidths of the
signal and idler fields differ accordingly. Such states are
interesting in some applications, e.g., in constructing heralded
single-photon sources.

\begin{figure}    
 \resizebox{0.7\hsize}{!}{\includegraphics{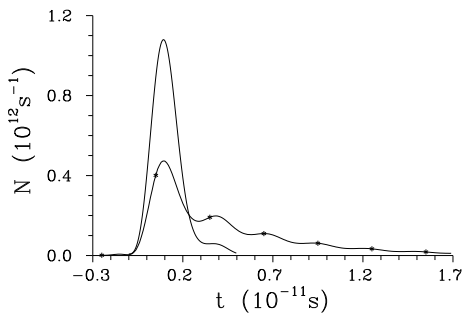}}
 \vspace{2mm}

 \caption{Photon fluxes of the signal ($ {\cal N}_s $, solid line) and idler
  ($ {\cal N}_i $, solid line with *) fields for a structure with
  $ N_{\rm elem} = 250 $ layers having two different transmission peaks in
  collinear geometry. The ratio of their intensity bandwidths equals cca 4.
  The structure is pumped by a pulse 250~fs long.}
\label{fig12}
\end{figure}

The two-photon spectral amplitude $ \phi(\omega_s,\omega_i) $ has
a cigar shape prolonged along the frequency of the field with a
larger spectral bandwidth (say the signal field). On the other
hand, the two-photon temporal amplitude $ {\cal A}(t_s,t_i) $ is
broken into several islands along the axis giving the idler-photon
detection time $ t_i $. Phase modulation of the two-photon
spectral amplitude $ \phi(\omega_s,\omega_i) $ with faster changes
along the frequency $ \omega_i $ and slower changes along the
frequency $ \omega_s $ is responsible for this behavior. Photon
fluxes of the signal and idler fields are then composed of several
peaks as documented in Fig.~\ref{fig12}. This feature is reflected
in an asymmetric dip in coincidence-count rate $ R_n^{\rm HOM} $
in Hong-Ou-Mandel interferometer, that also shows oscillations at
the difference of the central signal- and idler-field frequencies.
Asymmetry of the interference dip shown in Fig.~\ref{fig12} can be
related to the shape of spectral two-photon amplitude $ \phi $
\cite{PerinaJr1999}. Origin of these effects lies in delays caused
by the zig-zag movement of the generated photons inside the
structure.

\begin{figure}[tb]    
 \resizebox{0.7\hsize}{!}{\includegraphics{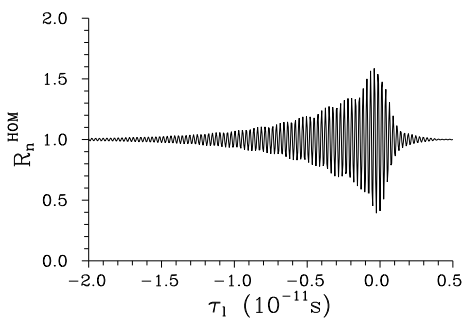}}
 \vspace{2mm}

 \caption{Coincidence-count rate $ R_n^{\rm HOM} $ in Hong-Ou-Mandel interferometer
  as a function of relative time delay $ \tau_l $
  for the two-photon state used in Fig.~\ref{fig12}.}
\label{fig13}
\end{figure}

\section{Conclusions}

Nonlinear random layered structures in which an optical analog of
Anderson localization occurs have been analyzed as a suitable
source of photon pairs with narrow spectral bandwidths and perfect
indistinguishability. Spectral bandwidths in the range from 1~nm
to 0.01~nm are available for different realizations of a random
structure. Photon pairs with the same signal- and idler-field
bandwidths as well as with considerably different bandwidths can
be generated. Random structures are flexible as for the generated
frequencies and emission angles. Two-photon states with coincident
frequencies and variable spectral bandwidth can be reached if
two-photon amplitudes of photon pairs generated into different
emission angles are superposed. Also two-photon states with
signal- and idler-field spectra composed of several peaks and
characterized by collective spectral mode functions are available.
All these states are very perspective for optical implementations
of many quantum information protocols. Possible implementation of
these sources into integrated optoelectronic circuits is a great
advantage.

\acknowledgments

The authors acknowledge support by projects IAA100100713 of GA AV
\v{C}R, COST 09026, 1M06002, and MSM6198959213 of the Czech
Ministry of Education. Also MIUR project II04C0E3F3 Collaborazioni
Interuniversitarie ed Internazionali tipologia C and support
coming from cooperation agreement between Palack\'{y} University
and University La Sapienza in Roma is acknowledged. The authors
thank fruitful discussions with A. Messina.

\end{document}